\documentclass{article}

\usepackage{arxiv}

\usepackage[utf8]{inputenc} 
\usepackage[T1]{fontenc}    
\usepackage{hyperref}       
\usepackage{url}            
\usepackage{booktabs}       
\usepackage{amsfonts}       
\usepackage{nicefrac}       
\usepackage{microtype}      
\usepackage{lipsum}
\usepackage{graphicx}

\usepackage{algorithmic}
\usepackage{algorithm}
\usepackage{listings}
\usepackage{seqsplit}
\usepackage{multirow}
\usepackage{colortbl}

\usepackage{authblk}

\title{Automatic Generation of a Cryptography Misuse Taxonomy Using Large Language Models}

\author[1,2,3]{Yang Zhang\thanks{The email addresses of Yang Zhang and Liang Cheng are \{zhangyang, chengliang\}@iscas.ac.cn.}}
\author[4]{Wenyi Ouyang\thanks{Both Yang Zhang and Wenyi Ouyang contributed equally to the paper.}}
\author[4]{Yi Zhang\thanks{The email addresses of Wenyi Ouyang and Yi Zhang are \{ouyangwenyi, zhangyi\}@ioccs.cn.}}
\author[1,2]{Liang Cheng\thanks{Yi Zhang and Liang Cheng are corresponding authors.}}
\author[5]{Chen Wu\thanks{wu.chen@microsoft.com}}
\author[1,2]{Wenxin Hu\thanks{huwenxin22@mails.ucas.ac.cn}}

\affil[1]{TCA Lab, Institute of Software, Chinese Academy of Sciences, Beijing, China}
\affil[2]{University of Chinese Academy of Sciences, Beijing, China}
\affil[3]{Zhongguancun Lab, Beijing, China}
\affil[4]{Institute of Cryptography and Cyberspace Security (Huangpu), Guangzhou, Guangdong, China}
\affil[5]{Microsoft Asia-Pacific Technology Company Ltd., Shanghai, China}

\begin{document}

\maketitle

\begin{abstract}
  The prevalence of cryptographic API misuse (CAM) is compromising the effectiveness of cryptography and in turn the security of modern systems and applications. Despite extensive efforts to develop CAM detection tools, these tools typically rely on a limited set of predefined rules from human-curated knowledge. This rigid, rule-based approach hinders adaptation to evolving CAM patterns in real practices.  

  We propose leveraging large language models (LLMs), trained on publicly available cryptography-related data, to automatically detect and classify CAMs in real-world code to address this limitation. Our method enables the development and continuous expansion of a CAM taxonomy, supporting developers and detection tools in tracking and understanding emerging CAM patterns. Specifically,  we develop an LLM-agnostic prompt engineering method to guide LLMs in detecting CAM instances from C/C++, Java, Python, and Go code, and then classifying them into a hierarchical taxonomy. 
  
  Using a data set of 3,492 real-world software programs, we demonstrate the effectiveness of our approach with mainstream LLMs, including GPT, Llama, Gemini, and Claude. It also allows us to quantitatively measure and compare the performance of these LLMs in analyzing CAM in realistic code.  Our evaluation produced a taxonomy with 279 base CAM categories, 36 of which are not addressed by existing taxonomies. To validate its practical value, we encode 11 newly identified CAM types into detection rules and integrate them into existing tools. Experiments show that such integration expands the tools' detection capabilities. 
\end{abstract}

\section{Introduction} \label{sec:intro}
Cryptographic algorithms are fundamental to the security of modern software, IT systems, and mobile technologies, ensuring secure communication, protecting sensitive data, and enforcing authentication and authorization mechanisms. However, misuse of cryptographic algorithms -- often manifested through incorrect or inappropriate API invocations -- can severely undermine the efficacy of these protections, jeopardizing the security and integrity of the systems they safeguard. Alarmingly, cryptographic misuse is prevalent; recent studies indicate that 88\% of Android applications contain at least one instance of cryptographic API misuse (CAM), exposing them to various security vulnerabilities~\cite{WWS+23}.

Detecting and mitigating CAM is therefore crucial to protecting software-driven systems. A significant body of research has focused on detecting CAM across diverse programming languages and hardware platforms. Techniques such as static analysis~\cite{RXA+19, KNR+17}, dynamic analysis~\cite{LGL+20}, and machine learning have been employed to enhance the detection of CAM. However, the scope of these detection methods — the types of CAM they can detect — is largely defined by their underlying taxonomies. These taxonomies consist of collections of patterns and rules that define cryptographic misuse, such as the six common misuse patterns proposed by Egele et al. in 2013~\cite{egele2013empirical}. While numerous extensions have been proposed since then, such as~\cite{muslukhov2018source, zhang2019cryptorex, wickert2021python, feichtner2018automated, li2014icryptotracer, wang2017nativespeaker}, these extensions have predominantly relied on manual encoding by human experts. 

The manual construction of CAM taxonomies constrains their comprehensiveness, limiting them to the expertise and awareness of human contributors. Although recent efforts attempt to synthesize knowledge from multiple sources to develop more systematic CAM classifications~\cite{ami2022crypto, afrose2022evaluation}, these approaches remain limited. For example, Ami et al.~\cite{ami2022crypto} identified nine CAM categories associated with insecure encryption modes such as \textsf{ECB} and \textsf{CBC}, yet overlooked other insecure modes, including \textsf{CFB}, \textsf{CTR}, and \textsf{XTS}, recognized by NIST standards~\cite{dworkin2001recommendation, saraf2014text}.

Additionally, manually curating taxonomies limits their ability to evolve continuously, as emerging CAM practices must first be identified and translated into detection rules by human experts. This delay hampers the capacity of existing taxonomies and associated detection tools to keep up with rapidly changing cryptographic practices and technological advancements.

Recent advances in generative large language models (LLMs) offer a potential solution to these limitations. State-of-the-art LLMs, such as ChatGPT and Gemini, are trained on extensive corpora that include cryptographic knowledge and software code, enabling them to process and analyze multi-modal information related to CAM. This capability positions LLMs as promising tools for autonomously developing and dynamically updating comprehensive CAM taxonomies from real-world code, thereby circumventing the limitations inherent to manual approaches.

In this paper we introduce an LLM-agnostic prompt engineering methodology that guides LLMs through a three-step process to detect, summarize, and classify CAM instances from realistic code written in C/C++, Java, Python, and Go. We prototype this approach in a tool named \textbf{CryMisTa} (\underline{Cry}ptographic \underline{Mis}use \underline{Ta}xonomy). CryMisTa demonstrates effectiveness with five mainstream LLMs -- ChatGPT 3.5, ChatGPT 4.0, Llama 3, Gemini 1.5, and Claude 3.5 -- allowing the development of comprehensive CAM taxonomies.  

Additionally, CryMisTa enables a quantitative comparison of LLMs on two key performance indicators affecting the taxonomy construction: (1) correctness and completeness in CAM detection, and (2) accuracy in categorizing CAM instances into coherent taxonomies based on their characteristics. Among the evaluated models, ChatGPT 4.0 and Claude 3.5 exhibit superior performance, under the same guidance of CryMisTa.

Applying CryMisTa to a dataset of 3,492 real-world programs, we generate a CAM taxonomy that surpasses existing taxonomies in comprehensiveness. It covers most CAM categories addressed by prior work and introduces 36 novel categories not previously identified. We foresee two primary benefits of such a comprehensive taxonomy. First, it can better facilitate human developers in understanding and organizing their knowledge of CAM, enabling them to write more secure code. 

Second, a more comprehensive CAM taxonomy with precise CAM descriptions and code examples can facilitate the enhancement of existing CAM detection techniques. To validate this utility of our taxonomy, we select 11 newly identified CAM categories and encode them as either detection rules compatible with tools such as CryScanner~\cite{CryScanner} or detection logic in Python. Experimental results show that incorporating the encoded rules allows CAM detection tools to identify CAM types that were previously undetectable.

In summary, this paper makes the following main contributions:
\begin{enumerate}
    \item We propose an LLM-agnostic prompt engineering methodology for constructing hierarchical CAM taxonomies from real-world code written in multiple programming languages. This approach is prototyped as the CryMisTa tool, and its efficacy is demonstrated using five state-of-the-art LLMs.
    \item We perform a quantitative evaluation of five widely used LLMs on critical performance indicators that impact the comprehensiveness and structure of CAM taxonomies, offering practical guidance on selecting LLMs in future CAM analysis tasks.
    \item Using CryMisTa on real-world code, we generate a comprehensive CAM taxonomy comprising 279 base categories, including 36 not addressed by existing taxonomies.
    \item We validate the utility of the CryMisTa-generated taxonomy by using the newly identified CAM categories to enhance existing CAM detection techniques, as confirmed by our experimental results.
\end{enumerate}

To promote transparency, reproducibility, and further advancement, we release our research artifacts publicly at \url{https://github.com/ufooooy/LLM-CMT}.

\section{Background And Related Work}\label{sec:background}

\subsection{Cryptography API Misuse Detection}
Most existing CAM detection technologies require the predefined specification of CAM types to be examined, typically in the form of coding patterns that encapsulate the essence of the misuse. The subject program's source code is then scanned and compared against these patterns to identify and locate instances of CAM.

Existing approaches collect the CAM categories by studying cryptographic algorithm documentation, academic literature, and other information sources. Egele et al.\cite{egele2013empirical} proposed six misuse patterns for Java cryptography libraries, which cover common misuses in symmetric key encryption, cryptography-based encryption, and random number generation. These patterns were adopted by BinSight\cite{muslukhov2018source}, CryptoREX\cite{zhang2019cryptorex}, LICMA\cite{wickert2021python}, Feichtner et al.\cite{feichtner2018automated}, iCryptoTracer~\cite{li2014icryptotracer}, and NativeSpeaker~\cite{wang2017nativespeaker}. 

Specifically, CryptoREX applied Egele's CAM patterns to detect CAM in IoT firmware, while LICMA applied them to detect CAMs in Python cryptography libraries and certain Java Cryptography Architecture (JCA) classes. FireBugs~\cite{singleton2021firebugs} selected five patterns from Egele's taxonomy to automatically detect and repair CAMs in applications using JCA. K-hunt~\cite{li2018k} proposed a dynamic detection tool that uses three of Egele's patterns to track cryptography key misuses.

With the advance of CAM detection research, researchers tend to expand misuse categories from multiple sources. For example, CryptoGuard~\cite{RXA+19} collected 16 rules from existing literature, based on which it created a 112-case benchmark {\sf cryptoAPI-bench} to evaluate CAM detection tools. Afrose et al.~\cite{afrose2022evaluation} designed ApacheCryptoAPI-Bench for Android apps and expanded their 16 CAM categories to 18 based on an extensive study on NIST guidelines and CAM-related blogs. CryLogger\cite{LGL+20} detects CAM in Android apps based on 16 misuse rules collected from both academic papers and NIST/IETF guidelines.

CogniCrypt$_{SAST}$~\cite{krugerCrySLExtensibleApproach2018} incorporated 23 misuse rules manually compiled from JCA documentation and introduced a specification language, CrySL, to facilitate the translation of such rules into constraints for CAM detection. Subsequently, RVSec~\cite{torresRuntimeVerificationCrypto2023} consolidated 22 rules from CogniCrypt$_{SAST}$, CryLogger, NIST's Juliet test suite\cite{juliet}, and OWASP's test suite\footnote{Owasp benchmark project, available at https://owasp.org/www-project-benchmark/}. These rules were then used by the JavaMOP tool~\cite{jinJavaMOPEfficientParametric2012} to detect CAM in Java and Android applications. As pointed out by its authors, the misuse rules of CryLogger cannot check temporal constraints, such as missing encryption steps. 

Sazzadur et al.~\cite{rahamanTheoryCodeIdentifying2022} formalized 25 CAM rules from CrySL and CryptoGuard as finite-state machines to facilitate the detection of 12 types of vulnerabilities in cryptographic libraries. CryptoGo\cite{li2022cryptogo} was the first CAM detector for Go cryptography libraries, which inherited 12 misuse rules from previous research. As indicated in its experiments, misuse rules for one programming language may not apply to others. 

The concept of a CAM taxonomy was first introduced in \cite{ami2022crypto}, where the taxonomy was manually constructed using CAM categories curated from top-tier conference papers (1999–2022), other academic literature, and industry guidelines. This taxonomy included 105 CAM categories, organized into 9 clusters. The categorization was guided by two principles: (1) the security objectives or properties violated by the misuse, and (2) the affected communication and computation stack, such as SSL or TLS. The manual construction of this taxonomy took approximately two person-months of effort.

CAM rules proposed in these studies were manually defined based on human expert knowledge. Translating them into a form actionable by CAM detection tools, such as machine-readable patterns and constraints, can be prone to errors. As indicated in \cite{chenPreciseReportingCryptographic2024}, errors in coding these rules are a major contributing factor to the false positives reported by CAM detection tools. More importantly, relying solely on human knowledge to continuously refine and expand CAM rules is challenging, especially in the face of the ever-evolving cybersecurity landscape.

To continuously improve the capability and accuracy of CAM detection tools, it is essential to develop an automated approach capable of constructing and updating a CAM taxonomy from realistic software development practices and cyber attacks. By leveraging the capability of cutting-edge LLMs, our approach enables the automatic development of an ever-growing CAM taxonomy from real-world code, which can serve as a dynamic guide for developers on understanding and avoiding CAMs in their daily tasks.

\subsection{LLMs and LLM-Based Taxonomy Development}
Modern LLMs, exemplified by the ChatGPT series, have been successfully applied to tasks of understanding and summarizing real-world code in different programming languages (e.g., \cite{WLG+23}) and text-based classification and categorization. Their effectiveness stems from extensive pre-training on a broad range of publicly available data, enabling them to accurately categorize text into predefined classes. For example, Rodrigues~\cite{rodrigues2023use} assessed ChatGPT's proficiency in classifying domain-specific terms according to upper-level ontologies, comparing its performance with expert classifications to gauge its accuracy and identify misclassifications.

In addition, LLMs excel in summarization tasks, generating concise and pertinent summaries of extensive documents. Their training regime includes tasks designed to emulate summarization, allowing them to produce summaries that preserve the original text's tone and context. Studies, such as those highlighted by Zhang~\cite{zhang2024benchmarking}, have shown that summaries generated by LLMs such as GPT models can match, and sometimes surpass, traditional methods in coherence and fidelity. In the news domain, LLM-generated summaries are comparable to those crafted by human experts.

General purpose LLMs have played a pivotal role in taxonomy development, assisting in structuring information into coherent systems of categories. Chen et al. \cite{chen2023prompting} was the first to use generative LLMs for taxonomy construction on data from the ACM Computing Classification System. It compared the prompting approach with fine-tuning methods, concluding that prompting outperforms fine-tuning but poses more challenges in post-processing. It formulated taxonomy generation as a text generation task and proposed several prompts for constructing taxonomies without explicit training. TacoPrompt~\cite{xu2023tacoprompt} was a multi-task prompt learning model designed for self-supervised taxonomy completion. Its core was a framework that integrated various tasks such as relation prediction and hierarchical classification, significantly improving overall performance. 

Subsequently, Moraes et al.\cite{moraes2024using} proposed an unsupervised method to utilize LLMs for the automatic construction and expansion of taxonomies. This approach combined topic modeling and keyword extraction techniques to create an initial topic taxonomy, which was then iteratively refined using LLMs. Zeng et al. \cite{zeng2024chain} proposed a "chain-of-layer" method to iteratively use LLMs for inducing taxonomy from limited examples. Gunn~\cite{gunn2024creating} leveraged GPT-4 to automatically develop a fine-grained entity type taxonomy on top of the UltraFine Entity Typing framework \cite{choi2018ultra}, in which ChatGPT demonstrated excellent language understanding capabilities, enabling it to parse complex texts, discern relationships, and propose hierarchical structures that experts can further refine. 

Our approach is agnostic to LLMs because its prompt-engineering method can work with any LLM sufficiently trained on cryptography-related information. Our approach distinguishes itself from other LLM-based taxonomy development research in that it targets the detection, summarization, and classification of CAMs in real-world code across different programming languages. Nonetheless, insights from prior research can be integrated into our approach to improve the construction of a taxonomy from the classified CAM categories.

\section{Method} \label{sec:method}
Intuitively speaking, generating a CAM taxonomy from a corpus of input code involves three phases.  
\begin{enumerate}
    \item {\bf Identification}: detects all CAM instances in the input code and summarizes the essential characteristics of the identified instances. This step should address the diversity of the input code, particularly when the corpus includes code written in different programming languages or targeted at different platforms. 
    \item {\bf Classification}: groups CAM instances sharing similar characteristics into categories. The summaries of CAM instances generated in the previous step should provide sufficient information to support accurate classification.
    \item {\bf Taxonomy Construction}: further groups the CAM categories into super-categories based on predefined criteria. This step organizes the CAM categories into a hierarchical structure by establishing relationships among categories. It is crucial that the predefined criteria align with the human cognitive process of cryptographic algorithms and their real-world applications, so that the resulting taxonomy can facilitate, rather than hamper, human comprehension.
 \end{enumerate}  
    
LLMs like ChatGPT have shown exceptional capabilities in code understanding, text summarization, and term classification tasks. Our preliminary experiments demonstrated that modern LLMs have been pre-trained with vast publicly available information related to cryptography, making them adept at detecting CAMs in real-world code. These advanced capabilities of LLMs can be leveraged to assist, and potentially automate, each step of the above taxonomy generation process.

However, using LLMs to construct CAM taxonomies has challenges:
\begin{itemize}
    \item {\bf Token Limitation}. Despite the continuous advancements of modern LLMs, their application in understanding and analyzing realistic code is constrained by token limits. Efficient mechanisms are needed to: (1) segment the input code as necessary to circumvent token limits, and (2) store and forward intermediate results when LLMs batch-process a large code corpus.
    \item {\bf Hallucination}. LLMs may sometimes generate outputs based on non-factual knowledge, a phenomenon known as {\it hallucination}. This can particularly affect the accuracy and precision of CAM detection. Thus, it is crucial to ensure that the LLMs use credible information sources for CAM identification, as suggested by previous research~\cite{MSV+24, TMJ+24}.
    \item {\bf Randomness}. Randomness can be introduced to LLMs during their design, training, and configuration. For instance, randomness may arise due to seed initialization~\cite{FWG20} and stochastic sampling methods such as temperature sampling~\cite{ZLL+24} and top-k sampling~\cite{VYP+23}. This randomness can lead to diversity and inconsistency when LLMs summarize CAM instances and classify CAM categories. Therefore, measures must be taken to control the impact of this randomness on the reproducibility, consistency, and coherence of the resulting taxonomies.
\end{itemize}

\begin{figure*}[hbtp]
    \centering
    \includegraphics[width=0.95\textwidth]{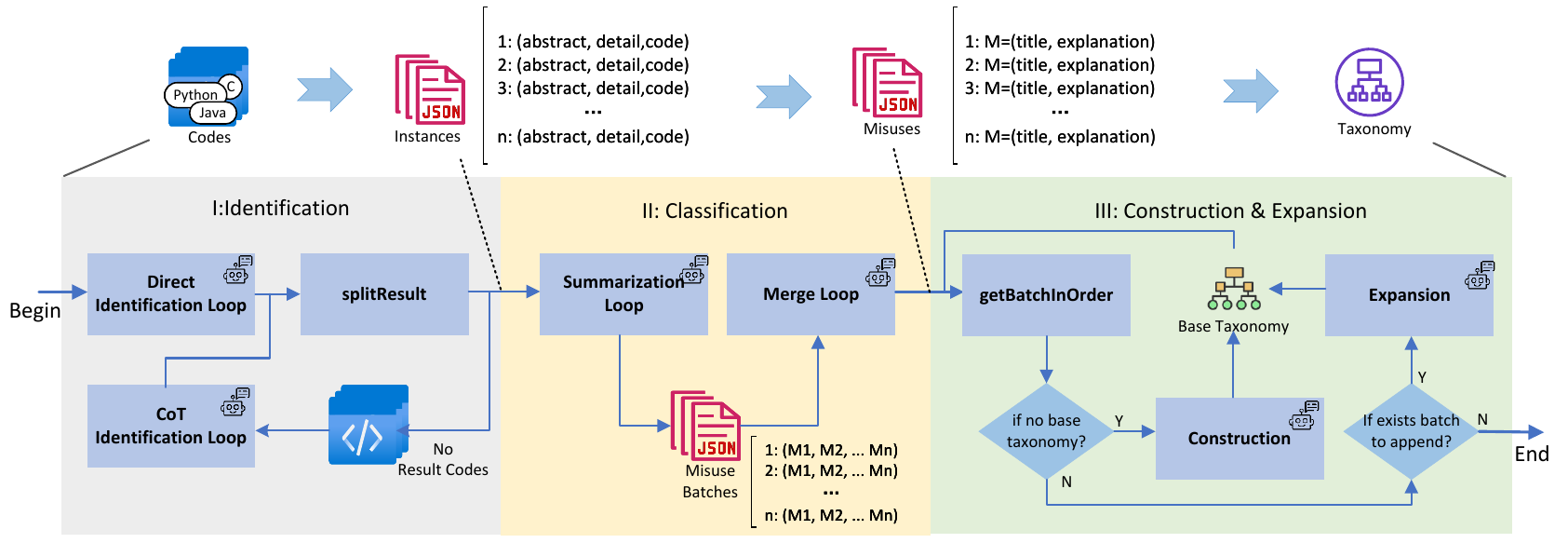}
    \caption{The Workflow of CryMisTa}
    \label{fig:pipeline}
\end{figure*}

To address the above challenges, we propose a work pipeline as illustrated in Figure~\ref{fig:pipeline} that integrates prompt engineering methods, customized batch-processing procedures, and clearly defined classification criteria to guide LLMs in automatically accomplishing each phase of the CAM taxonomy construction process. This work pipeline, in alignment with the intuitive construction process, consists of three phases: Identification, Classification, and Taxonomy Construction. Notably, this work pipeline theoretically can work with any general-purpose LLM. 

\subsection{The Identification Phase}
\begin{figure}[hbtp]
    \centering
    \includegraphics[width=0.45\textwidth]{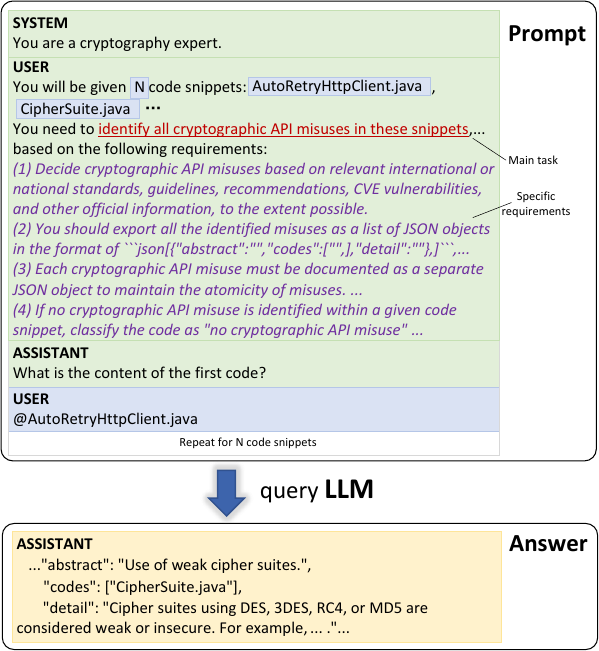}
    \caption{An Example Inquiry-Response for Direct CAM Identification}
    \label{fig:direct_id}
   \vspace{-6pt}
\end{figure}

Our approach instructs LLMs to report a CAM \textit{Instance} as a triplet $\langle abstract, detail, code\rangle$, where $abstract$ is a succinct description of the instance, $detail$ is a more elaborated version of $abstract$, and $code$ is the identifier of the program in which the CAM is detected. This structure helps to reduce the randomness of LLMs in reporting CAM instances and normalizes the level of information to be included in such reporting to facilitate subsequent phases. For example, the lower part of Figure~\ref{fig:direct_id} provides an example CAM instance that is detected from {\sf CipherSuite.java}, in which the abstract \textit{Use of weak cipher suites} summarizes the nature of the misuse, and the detail section provides a more detailed digest of it.  

Given a piece of code $c$, an LLM may detect zero, one, or multiple CAM instances in $c$. Let $count$ be the number of CAM instances detected in $c$. In response to the identification prompt (shown in the top part of Figure~\ref{fig:direct_id})), the LLM may return one of the following: 1) one instance $I$ with an abstract of ``No misuse" if $count = 0$; 2) one instance $I$ summarizing the detected instance if $count=1$; or 3) $count$ instances, where each instance summarizes the corresponding CAM if $count >1$. 
   
The prompt depicted in Figure~\ref{fig:direct_id} imposes four requirements to normalize the LLM's identification and reporting of CAM instances. Requirement (1) instructs the LLM to utilize credible information sources, such as standards, CVE vulnerabilities, or other official documentation, to determine whether the input code contains CAM. Requirement (2) instructs it to report all identified CAM instances separately. Requirements (3) and (4) guide the LLM to report CAM instances in a structured format.  

One noteworthy aspect of this CAM detection prompt is its semi-structured design: the static part (highlighted in green) provides static yet specific instructions to the LLM on how to accomplish the task at hand for multiple entities of the same kind (e.g., CAM instances or categories), while the dynamic part (marked in blue) can be customized for the entity currently being processed (e.g., intermediate results generated so far). This approach facilitates the development of batch-processing scripts for a large corpus of realistic code and is therefore adopted by all other prompts in our workflow.

To overcome the token limits of LLMs, the Identification Phase utilizes two batch-processing loops: the Direction Identification Loop and the Chain of Thought (CoT) Identification Loop. During the Direction Identification Loop, the zero-shot prompt depicted in Figure~\ref{fig:direct_id}) is repeatedly used to instruct the LLM to detect CAM instances in batches of input code, storing the returned instances in a JSON file.

However, our preliminary experiments indicated that LLMs, particularly ChatGPT 3.5, sometimes struggled with effectively applying their cryptography knowledge in CAM detection. To address this issue, we implement a filtering process (represented by the $splitResult$ box in Figure~\ref{fig:pipeline}), which identifies code segments where the Direction Identification Loop cannot conclusively determine the presence of CAM and forwards them to the CoT Identification Loop for more comprehensive analysis. 

The CoT Identification Loop employs a series of prompts to guide the LLM through a chain-of-thought process for identifying CAM instances. The CoT prompts~\footnote{Detailed in Appendix A.} first instruct the LLM to identify elements of the input code related to cryptography, then prompt it to search for knowledge of related misuse practices, and lastly instruct it to apply such knowledge to analyze the input code. When one or more CAM instances are detected, the LLM summarizes and reports these instances in a structured format, akin to the Direct Identification Loop. 

With the assistance of the CoT Identification Loop, ChatGPT-3.5 encountered no situations where it could not decide if the input code contained CAM in our experiments (see Section~\ref{sec:evaluate} for details).

\subsection{The Classification Phase}
The inherent randomness of LLMs can introduce variability to the summaries generated for the detected CAM instances, which may complicate the accurate classification of CAM instances. To address this issue, we standardize a CAM category as a tuple $\langle title, explanation\rangle$, where $title$ is a concise summary of the type of CAM and $explanation$ provides a detailed description of that type of misuse, potentially with code examples. 

With this standardized notion of a \textit{CAM Category}, classifying CAM instances involves two primary tasks: (1) identifying CAM instances of the same nature and summarizing them into CAM categories, and (2) merging similar or identical CAM categories to eliminate redundancy. The Classification Phase in our approach employs two batch-processing loops — the Summarization Loop and the Merging Loop — to accomplish these tasks, respectively.

\subsubsection{The Summarization Loop}
\begin{figure}[hbtp]
    \centering
    \includegraphics[width=0.45\textwidth]{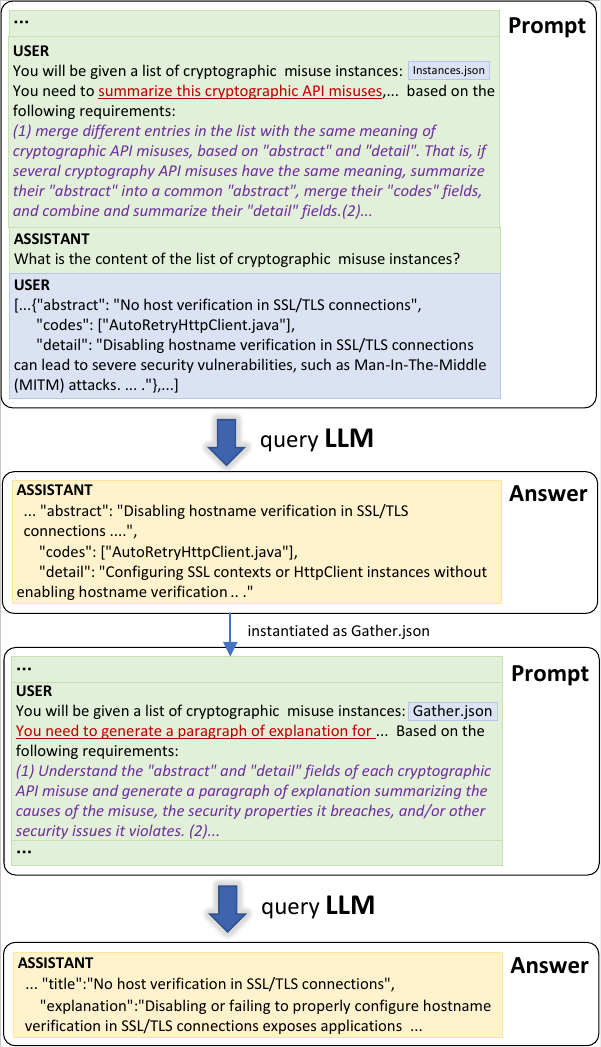}
    \caption{An Example of Misuse Summarization}
    \label{fig:misuse_sum}
\end{figure}

The Summarization Loop employs the prompts illustrated in Figure~\ref{fig:misuse_sum} to repeatedly instruct the LLM to identify CAM instances of the same nature and generate CAM categories that summarize their commonalities. Specifically, the top prompt in Figure~\ref{fig:misuse_sum} instructs the LLM to compare the misuse summaries based on their $abstract$ and $detail$ fields. It then aggregates those sharing the same misuse into a \textit{merged} instance, where the $abstract$ provides a common title for all involved instances and its $detail$s merges the code identifiers and concatenates the $detail$s of these instances (Requirement 1). The bottom prompt, on the other hand, instructs the LLM to understand the $abstract$ and $detail$ of the \textit{merged} instances, and creates a $title$ and $explanation$ for each CAM category that accurately summarize these instances. 

\subsubsection{The Merging Loop}
As the Summarization Loop categorizes CAM instances in batches, redundant CAM categories may be created for the same type of CAM instances, if they occur in different batches. The Merging Loop is designed to consolidate identical or similar CAM categories: it repeatedly prompts the LLM to compare CAM categories generated in different batches to determine whether they essentially represent the same type of CAM and retain only one out of all identical CAM categories. This ensures that each CAM category at the completion of the Merging Loop represents a distinct CAM type.

\renewcommand{\algorithmicrequire}{\textbf{Input:}}  
\renewcommand{\algorithmicensure}{\textbf{Output:}} 

\subsection{The Taxonomy Construction Phase}
The output of the Classification Phase is a set of CAM categories. Constructing a hierarchical taxonomy for these CAM categories involves introducing high-level CAM categories to summarize the common attributes of their sub-categories, so as to organize all CAM categories into a multi-layer structure to facilitate human comprehension. The high-level CAM categories introduced during this process differ from the original CAM categories, as the latter are derived directly from realistic code. We refer to the CAM categories derived from realistic code as $base categories$ and those newly introduced as $abstract categories$. 

Formally, a CAM taxonomy $T$ is a tuple $\langle N, E \rangle$, where $N$ is a set of CAM categories (base or abstract) and $E: N\times N$ is the set of edges (or relations) among CAM categories. An edge $(p, c)\in E$ indicates that category $p$ is the parent of category $c$, summarizing the common characteristics of $c$ and its sibling categories.  

The CAM taxonomy construction process starts with the set of base CAM categories and $E=\emptyset$. As the construction proceeds, abstract categories are introduced to summarize their children categories (base or abstract), thereby expanding both $N$ and $E$ of the taxonomy. 

As previously discussed, there are multiple ways to classify base CAM categories into abstract categories, depending on the attributes used to find their commonalities. For instance, the ``Insufficient Salt Length" category can be classified under ``Insecure Use of Cryptographic Parameters" if the root cause of misuse is the classification attribute. Alternatively, it can be categorized under ``Salt Misuses" if the affected cryptographic system component is the classification attribute.

The selection of appropriate attributes for classifying CAM categories should align with human cognitive processes and consider the taxonomy's intended application in realistic contexts. For instance, if the taxonomy is intended to guide the detection and resolution of CAM or assist developers in avoiding CAM during software development, classifying CAM categories based on their root causes may be more effective. Conversely, if the primary goal is to assess the severity of security vulnerabilities and design appropriate security controls, it may be more appropriate to classify CAM categories based on the affected cryptographic system components or security properties.

The current version of our approach employs a set of keywords from \cite{ami2022crypto} to guide the LLM in classifying CAM categories. These keywords, listed in Table~\ref{tab:dim}, describe the general root causes of CAM. However, our approach can be easily adapted to utilize alternative keyword sets, such as those focused on cryptographic system components, to tailor the classification for other use cases. Future work will explore how the choice of keywords impacts the utility of the resulting taxonomies.

\begin{figure}[hbtp]
    \centering
    \includegraphics[width=0.45\textwidth]{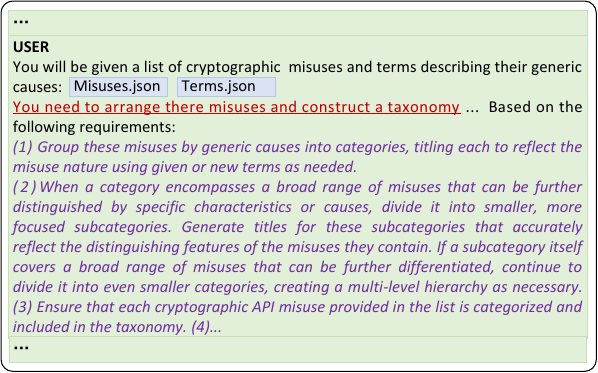}
    \caption{An Example of the Construction Prompt}
    \label{fig:constr_prmpt}
   \vspace{-6pt}
\end{figure}

\begin{figure}[hbtp]
    \centering
    \includegraphics[width=0.45\textwidth]{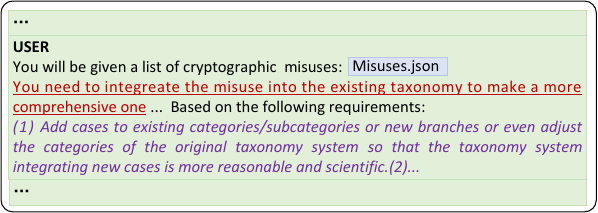}
    \caption{An Example of the Expansion Prompt}
    \label{fig:app_prmpt}
\end{figure}

As in earlier phases, taxonomy construction proceeds through batch processing of CAM categories to accommodate the token limits of the LLM. The process begins with constructing a baseline taxonomy, where the LLM classifies CAM categories from the initial batch using the provided keywords. It then continues by integrating CAM categories from subsequent batches into the baseline taxonomy. To support these tasks, we developed the Construction Prompt (shown in Figure~\ref{fig:constr_prmpt}) and the Expansion Prompt (illustrated in Figure~\ref{fig:app_prmpt}), which guide the LLM in building and expanding the taxonomy, respectively.

\textbf{The Construction Prompt.}
The Construction Prompt, shown in Figure~\ref{fig:constr_prmpt}, guides the LLM to refine a baseline taxonomy structure from a subset of the CAM category set generated in the Classification Phase, based on keywords (stored in ``{\sf Terms.json}"). This refinement may involve decomposing existing abstract categories or introducing new ones, as specified in Requirement 2. Importantly, the prompt also instructs the LLM to generate new keywords if necessary, ensuring they align with the existing ones (Requirement 1). In essence, the prompt provides the LLM with the original keywords as few-shot examples, enabling it to classify CAM categories more effectively and intelligently.

\textbf{The Expansion Prompt.}
Expanding the baseline taxonomy with new CAM categories follows a process analogous to the initial taxonomy construction. When introducing a new CAM category, the LLM is instructed to evaluate its alignment with existing categories in the taxonomy. If no alignment is found, it is instructed to create new connections or abstract categories to integrate the new category into the existing taxonomy structure (as specified in Requirement 1 of Figure~\ref{fig:app_prmpt}). Furthermore, the newly added categories must conform to the established structural format to ensure consistency across the taxonomy, as dictated by additional requirements in the Expansion Prompt (not shown in Figure~\ref{fig:app_prmpt} for brevity).

\begin{table}[hbtp]
  \renewcommand{\arraystretch}{1.3}
  \caption{Keywords for Classifying CAM categories}
  \label{tab:dim}
  \begin{tabular}{p{0.7\textwidth}}
  \toprule
  Keywords \\
  \midrule
 Compromising Client and Server Secrecy; Compromising Secrecy of Cipher Text; Compromising Randomness; Compromising Secret Keys; Compromising Integrity through Improper Checksum Use; Compromising Non-Repudiation; Compromising Communication Secrecy with Intended Receiver\\
  \bottomrule
  \end{tabular}
  \end{table}

\subsection{Prototyping}\label{sec:implemt}
We implemented our approach as an LLM-agnostic tool, CryMisTa, consisting of 1,317 lines of Python code. CryMisTa implements: (1) the workflow pipeline illustrated in Figure~\ref{fig:pipeline}, (2) the batching processes described in this section, and (3) basic static analysis and code-slicing functionalities to partition large programs into segments related to cryptographic API usage, ensuring compliance with the LLM token limits (currently set at 4,095).
\section{implementation}\label{sec:implemt}
Even though our approach is agnostic to any generative LLMs that are capable of understanding and summarizing software code, we chose the {\sf gpt-4-turbo-preview} model to implement a prototype system called CryMisTa. This model has a maximum token count of 4095, with the temperature value set to 0 to ensure more consistent and deterministic outputs\cite{openai-faq}, and the {\sf top\_p, frequency\_penalty}, and {\sf presence\_penalty} parameters set to 1, 0, and 0 respectively, aligning with the default settings. All the CMT stages described above are implemented in Python, with 1,317 lines of code in total. 
\section{Evaluation}\label{sec:evaluate}
We conducted a series of experiments to evaluate the effectiveness of CryMisTa and the usefulness of the CAM taxonomies it generates. To this end, we selected five widely used general-purpose LLMs as candidates: GPT-4 (gpt4-turbo-preview), GPT-3.5 (gpt3.5-turbo-preview), Llama-3 (llama3.1-405b-instruct), Gemini-1.5 (gemini-1.5-flash), and Claude-3.5 (claude-3.5-sonnet-20240620). During the evaluation, identical configurations were applied to all candidate LLMs to ensure a fair comparison. Specifically, the temperature, token limit, top\_p, frequency\_penalty, and presence\_penalty were set to 0, 4095, 1, 0, and 0, respectively. Notably, setting the temperature to 0 minimized randomness in the CAM instance summaries generated by the LLMs.

The experiments aimed to address the following research questions:
\begin{itemize}
	\item[RQ1:] Can CryMisTa operate with the candidate LLMs, and how well do these LLMs perform in the key tasks of constructing CAM taxonomies? (Effectiveness)
	\item[RQ2:] How does the quality of the taxonomy generated by CryMisTa compare to existing CAM taxonomies regarding comprehensiveness? (Advantage)
	\item[RQ3:] How effective is the taxonomy generated by CryMisTa in supporting the expansion of existing CAM detection techniques? (Usefulness) 
\end{itemize}

\subsection{Experiment Dataset}
We constructed a dataset of 3,492 software programs, denoted as $D$, sourced from two origins: (1) 457 programs from previous research, including 18 from CryptoGo, 196 from CryptoAPI-Bench, 40 from MASC, and 203 from LICMA\footnote{CryptoRex does not publicly release its code or dataset, so it was excluded from further evaluation.}; and (2) 3,035 real-world programs written in C/C++, Python, Java, and Go, crawled from GitHub repositories. The GitHub repositories selected for crawling met two criteria to ensure relevance and credibility: (1) the repository description contained keywords related to cryptography, such as \textsf{crypto}, \textsf{javax.crypto}, and \textsf{OpenSSL}; and (2) the repository had more than 100 stars and commits.

To establish a ``{\it ground truth}'' for evaluation, we selected a subset of $D$ and manually analyzed the use of cryptographic algorithms. This subset, denoted as $D_{sub}$, comprises 417 programs from prior research and 299 randomly selected from web-crawled programs, totaling 716 programs. Regarding programming languages, $D_{sub}$ includes 245 programs in Python, 231 in Java, 120 in C/C++, and 120 in Go. 

Two authors, both with substantial expertise in cryptography, independently reviewed $D_{sub}$ for CAM instances. Disagreements between them were discussed and resolved collectively by all authors. Our manual analysis identified 975 CAM instances across 540 programs in $D_{sub}$. Additionally, 118 programs were found to use cryptographic APIs correctly, while the remaining 58 were unrelated to cryptography.

\subsection{RQ1: Effectiveness}
In our initial experiments, CryMisTa successfully guided all candidate LLMs to construct hierarchical taxonomies from $D_{sub}$. To quantify their performance, we designed two experiments to evaluate their effectiveness in two key tasks: detecting CAM instances from input code and classifying CAM instance summaries into appropriate categories.

\subsubsection{Correcteness and Completeness of CAM Detection}
The first experiment assessed the correctness and completeness of the candidate LLMs in detecting CAM, where $D_{sub}$ was provided to the LLMs and the CAM instances reported by the LLMs were compared with those labeled by the authors.

\textbf{Correctness}
As noted earlier, $D_{sub}$ consists of 540 programs containing CAM (i.e., positive samples) and 176 programs free of CAM (i.e., negative samples). Table~\ref{tab:correct} lists the accuracy, precision, and recall of the LLMs in determining whether the input programs contain CAM. As shown in Table~\ref{tab:correct}, except for GPT-3.5, all candidate LLMs demonstrated consistently strong performance. For instance, GPT-4 correctly classified 527 out of 540 CAM-containing programs and 121 out of 176 programs without CAM. Among the models, GPT-4 demonstrated the best detection performance, by achieving the highest F1 score of 0.944. Due to its obvious disadvantage against the others, we excluded GPT-3.5 from subsequent experiments.

\begin{table*}[h!]
    \caption{Model Performance Comparison}
    \label{tab:correct}
    \begin{tabular}{lccccc}
        \toprule
        & GPT-4 & GPT-3.5 & Llama-3 & Gemini-1.5 & Claude-3.5 \\
        \midrule
        Accuracy & 0.91 (648/716) & 0.78 (560/716) & 0.88 (631/716) & 0.90 (644/716) & 0.90 (647/716) \\
        Precision & 0.91 (527/582) & 0.98 (392/400) & 0.89 (521/587) & 0.95 (496/526) & 0.92 (518/565) \\
        Recall & 0.98 (527/540) & 0.73 (392/540) & 0.96 (521/540) & 0.92 (497/540) & 0.96 (518/540) \\
        F1 & 0.944 & 0.790 & 0.924 & 0.935 & 0.940 \\
        \bottomrule
    \end{tabular}
\end{table*}

\textbf{Completeness}
It is possible that multiple CAM instances might exist in one single program. For example, the 540 CAM-containing programs in $D_{sub}$ contained 975 CAM instances, according to our manual analysis. It is hence crucial for LLMs to detect all CAM instances within a program (i.e., detection completeness) to ensure comprehensive contributions to the resulting taxonomy.

In the experiment, GPT-4 detected 1,470 CAM instances in $D_{sub}$, Llama-3 reported 1,544 instances, Gemini-1.5 identified 936, and Claude-3.5 found 1,654. Analysis of these results revealed two common patterns across all LLMs: (1) LLMs generally reported CAM instances at a more granular level than human reviewers, leading to a higher count of reported instances, and (2) they exhibited conservative tendencies, often issuing `warnings' rather than definitively labeling instances as CAM. For example, GPT-4 flagged a potential key management issue in {\sf CryptoLibAESCBCLV2.py}\footnote{The source code of {\sf CryptoLibAESCBCLV2.py} and its analysis results are available in our GitHub repository.}, with the statement: \textit{The absence of key management or validation mechanisms in the provided code may lead to the use of weak or inadequate keys for encryption}, instead of classifying it outright as a CAM instance. 

To quantify the detection completeness, we manually mapped the CAM instances reported by GPT-4 — the model with the best detection correctness — to those labeled by the authors. Reported instances that were identical to or subsets of some labeled instances were considered correct. Additionally, warnings were deemed correct, if the information and recommendations they conveyed were accurate. 

Our analysis found that GPT-4 falsely reported 24 CAM instances and missed 107 labeled instances. This corresponds to a False Negative Rate of 11.0\% (107/975) and a False Discovery Rate of 1.6\% (24/1470), which we deem acceptable. Note that we consider relatively high False Negative Rates as acceptable because we anticipate that as the number of input software programs increases, the (types of) CAM instances initially overlooked will eventually be captured by the LLMs within a larger and more diverse set of programs.  

After analyzing the results, we attributed GPT-4 missing 107 instances to two primary reasons. First, it may prioritize reporting the primary misuse instances detected in a program but overlook the others. For example, in the case of {\sf RC4.cpp}, it identified the ``Use of weak encryption algorithm (RC4)" misuse at lines 17-20 but missed ``Use of a hard-coded key" at line 8. Second, it may not have the relevant cryptography knowledge pre-trained or may not have correctly utilized such knowledge. For example, it missed reporting the use of the insecure {\sf CFB8} cipher mode or the insecure {\sf CAST5} algorithm. 

The false positives reported by GPT-4, on the other hand, fall into two main types: ``Hardcoded cryptographic algorithm (names)" and ``Insecure secret lengths, including key, IV, and salts". We speculate that these false positives were probably because GPT-4 relied on general cryptographic knowledge to detect CAM rather than performing an in-depth analysis of cryptographic API usage in the code.

The detection completeness of other LLMs can be calculated through the same manual analysis, which we leave for future work.

\begin{figure}[!t]
    \centering
    \includegraphics[width=\columnwidth]{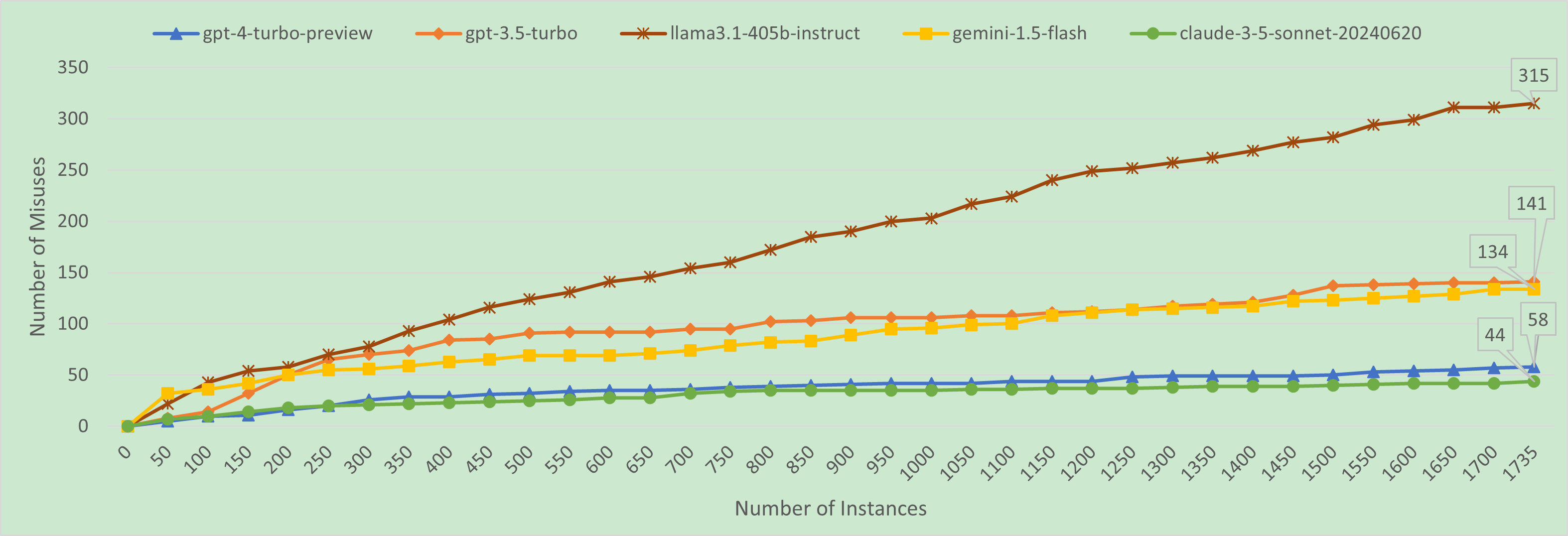}
    \caption{Classifying CAM Instances by LLMs}
    \vspace{-6pt}
    \label{fig:categories}
\end{figure}

\subsubsection{CAM Categorization}
Another key factor influencing the quality of the resulting taxonomies is the ability of LLMs to classify the detected CAM instances into appropriate categories with CryMisTa's guidance. This process involves understanding the summaries of CAM instances and performing necessary categorization actions, which may include creating new categories. Our second experiment quantitatively evaluates the classification performance of the candidate LLMs, where all LLMs were instructed by CryMisTa to classify summaries generated by GPT-4 in last experiment. A total of 1,735 summaries were generated from the $D_{sub}$ by GPT-4 , consisting of 1,470 summaries for CAM instances or warnings and 265 `no misuse' summaries.\footnote{GPT-4 generates a `no misuse' summary for each correct use of cryptographic APIs or for the entire program if no cryptography is involved.} We expect this approach can mitigate the influence of randomness in CAM instance summaries produced by different LLMs.  
Figure~\ref{fig:categories} illustrates the evolution of base CAM category numbers introduced by LLMs as more CAM instance summaries were provided. As shown in Figure~\ref{fig:categories}, GPT-4 and Claude-3.5 classified the input CAM instance summaries into the fewest categories (44 and 58, respectively), whereas Llama-3 produced the highest number of categories (315). With an increasing number of input summaries, the number of base CAM categories generated by Llama-3 rose rapidly, while GPT-4 and Claude-3.5 exhibited the slowest growth rates. In addition, Gemini-1.5 and GPT-3.5 demonstrated comparable performance, producing a moderate number of categories with a moderate growth rate.

These results suggest that GPT-4 and Claude-3.5 exhibit the strongest inductive capabilities, effectively grouping similar misuses into broader categories while demonstrating greater conservatism in introducing new categories for related or similar misuses. In contrast, Llama-3 tends to classify misuses in a more fine-grained manner. We believe that conservative models are more likely to emphasize commonalities among misuses, thereby minimizing unnecessary category expansion and contributing to a more stable and consistent classification system. By generating fewer categories, these models are less prone to misclassify uncertain or ambiguous misuse summaries, enhancing the readability and usability of the resulting taxonomies, especially for human users.

\begin{figure*}[!t]
    \centering
    \includegraphics[width=0.8\textwidth]{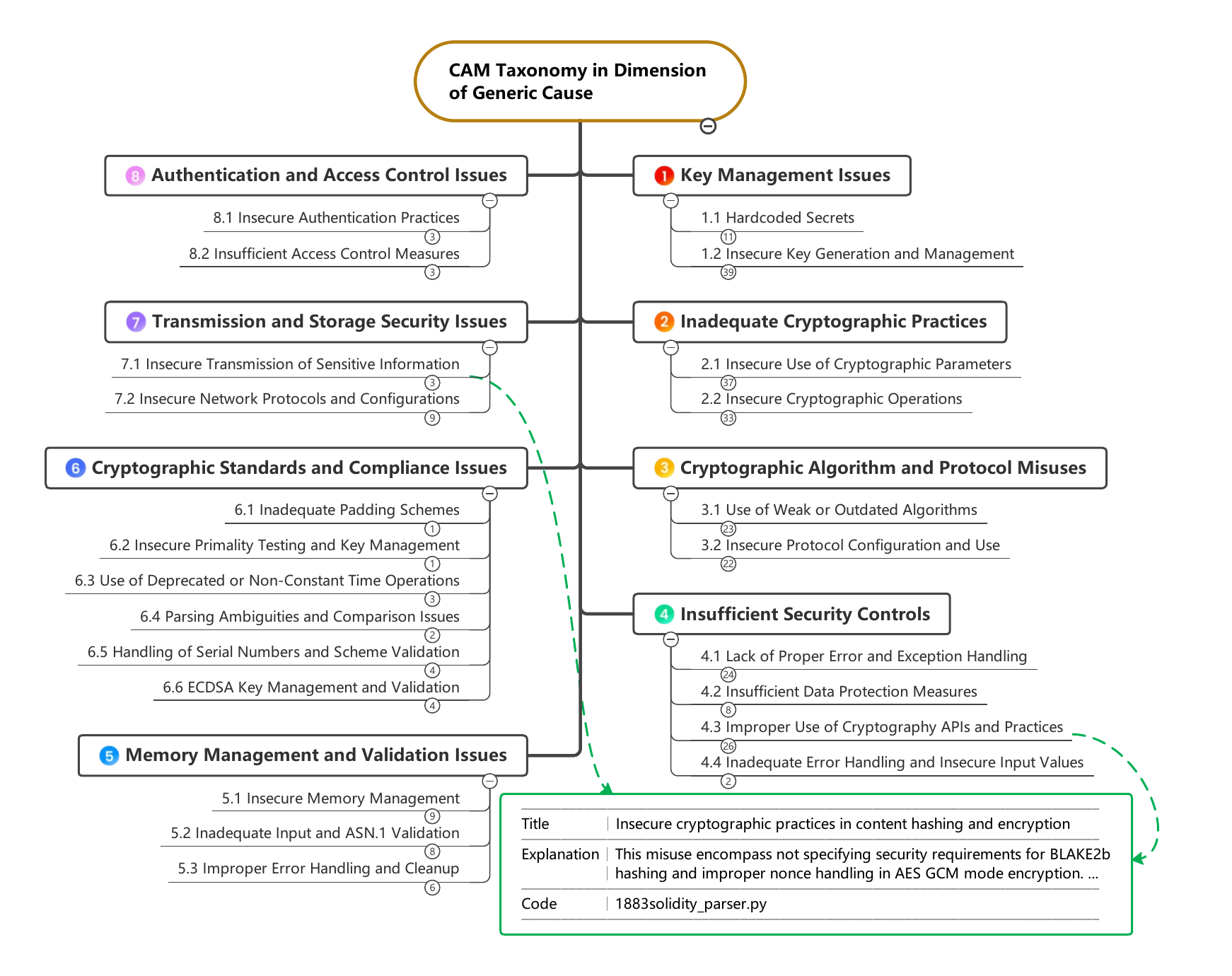}
    \caption{Top Levels of the CAM Taxonomy Generated by GPT-4 under the Guidance of CryMisTa}
    \vspace{-6pt}
    \label{fig:txmy_example}
\end{figure*}

In summary, our experiments concluded that all candidate LLMs, except GPT-3.5, demonstrated acceptable correctness in detecting CAM from realistic code and with GPT-4 achieved acceptable completeness in identifying all CAM instances. The candidate LLMs exhibited significant differences in classifying CAM instances. These results led us to conclude that GPT-4 achieved the best overall performance. Hence, our subsequent analysis focused on the taxonomy produced by GPT-4 for $D$, the top two levels of which are illustrated in Figure~\ref{fig:txmy_example}. However, we make the taxonomies that the other LLMs generated for $D_{sub}$ available in our GitHub repository, to highlight that CryMisTa can work with these models as well.

\subsection{RQ2: Advantage}\label{sub:advantage}
The taxonomy generated by GPT-4 for $D$ forms a three-layer hierarchy, consisting of 8 top-level categories, 23 second-level categories, and 279 base categories. Notably, the input order of programs in $D$ had minimal to no impact on the final structure of the taxonomy. We provided GPT-4 with $D$ five times, each with a different input sequence, and all resulting taxonomies consistently contained 279 base categories and 8 top-level categories. This consistency suggests that, given a sufficiently large input volume, CryMisTa enables GPT-4 — and potentially other LLMs — to reliably identify, summarize, and categorize CAM instances in a reproducible manner.

 We compared the taxonomy constructed by GPT-4 with existing taxonomies presented in CryptoRex, CryptoGo, CryptoAPI-Bench, MASC~\cite{amiMASCToolMutationBased2023}, and LICMA, w.r.t. the comprehensiveness of misuse categories. To ensure a fair comparison, we only considered the base CAM categories in the taxonomy generated by GPT-4.
 
Table~\ref{tab:txmy_cmp} presents the comparison results. The numbers in parentheses in the first column indicate the number of misuse categories included in each of the taxonomies under comparison. The column labeled ``CryMisTa $\cap$ X" shows the number of misuse categories that overlap between the GPT-4 taxonomy and existing taxonomies. The column labeled ``X-CryMisTa" reports the number of misuse categories present in the compared taxonomy but absent from the GPT-4 taxonomy. The rightmost column of Table~\ref{tab:txmy_cmp} lists the number of base CAM categories identified by GPT-4 that are not covered by any of the existing taxonomies.
 
\begin{table}
	\renewcommand{\arraystretch}{1.3}
	\caption{Comparison between the Taxonomy generated by CryMisTa and Existing Taxonomies}
	\label{tab:txmy_cmp}
	\begin{tabular}{lccc}
		\toprule
            {\bf Taxonomy}  & {\bf CryMisTa $\cap$ X} & {\bf X-CryMisTa}  & {\bf CryMisTa-X}\\
		\midrule
		CryptoRex(6)   & 6 & 0 &  85 \\ 
		CryptoGO(12)  & 12 & 0 &   44  \\
		CryptoAPI-Bench(16) & 16 & 0 &  36   \\
		MASC(105)  & 93 & 12 &  36 \\
    \bottomrule
	\end{tabular}
 \end{table}

As shown in Table~\ref{tab:txmy_cmp}, all misuse categories included in CryptoRex, CryptoGO, and CryptoAPI-Bench are fully covered by the GPT-4 taxonomy. However, the GPT-4 taxonomy does not account for twelve (12) misuse categories in MASC. Notably, for 10 of these 12 categories, we found no corresponding code samples in MASC's open-source dataset. MASC itself did not identify any code samples under these 10 categories when being applied to its open-source dataset.  The two remaining categories missed by the GPT-4 taxonomy — ``Constant password for Android keystore" and ``Android Webview incorrect certificate verification" — are specific to Android apps. They were omitted because $D$ did not include Android app code. When presented with a sample from MASC's dataset that exemplifies the latter category, CryMisTa guided GPT-4 to correctly identify and summarize it. This comparison confirms that CryMisTa can effectively guide LLMs to generate more comprehensive taxonomies when supplied with sufficiently large and diverse datasets.

We engaged external cryptography experts to independently review the 36 previously unaddressed misuse categories identified in the GPT-4 taxonomy. Their evaluation confirmed that these categories represent legitimate instances of insecure cryptographic algorithm usage, potentially leading to security vulnerabilities. This further highlights the strength of CryMisTa in guiding LLMs to generate more comprehensive taxonomies from realistic code. 

We briefly summarize these categories in the following areas and more details can be found in our GitHub repository. 

\subsubsection{Insecure use of specific cryptographic algorithms}
Three categories are related to the insecure use of specific cryptographic algorithms, involving hash function {\sf Keccak256}, symmetric encryption algorithm {\sf RC5}, and random number generation algorithm {\sf adler32}. Take {\sf Keccak256} for example. The ``Insecure Hashing Operation and Insecure Hash Function Usage" category captures using {\sf Keccak256} without proper context or data handling. Specifically, early versions of the {\sf Keccak} padding scheme and parameter settings used the outdated {\sf sha3.NewLegacyKeccak256()}, instead of the standard SHA-3 {\sf sha3.New256()} recommended by NIST FIPS 202~\cite{sha3}. 

The GPT-4 taxonomy includes a category for insufficient iterations in \textsf{PBKDF2}. As recommended iteration counts for \textsf{PBKDF2} have increased over time (e.g., OWASP advised 600,000 iterations for \textsf{PBKDF2-HMAC-SHA256} in 2023~\cite{pbkdf2}), existing taxonomies and CAM detection tools depending on them (such as LICMA) still rely on the outdated recommendation of 1,000 iterations. This discrepancy can result in insecure usage of \textsf{PBKDF2} being overlooked.
	
\subsubsection{Insecure encryption modes}
Seven categories concern insecure encryption modes not covered by existing taxonomies. For instance, ``Inappropriate GCM tag length" points out that, when using the {\sf GCM} encryption mode, the {\sf GCM} tag length is set in a way that does not follow recommendations in NIST SP 800-38D~\cite{sp800-38d}, thereby compromising data integrity and authenticity. ``Insecure tweak calculation in XTS mode encryption" indicates that a lack of sufficient randomness and unpredictability of the tweak value in the {\sf XTS} mode, which violates recommendation in \cite{sp800-38e, CRESTS}, can lead to security vulnerabilities\footnote{CryptoGo is the only taxonomy that points out the misuse of using non-unique IVs in the {\sf XTS} mode, but it did not mention misuse related to setting the tweak value.}.  

The remaining categories address (1) using inappropriate or unsupported IV length for the {\sf GCM} mode, (2) insecure IV length validation for the {\sf GCM} mode, (3) using inappropriate randomness sources (e.g., the secret key) in IV generation for AES encryption in the {\sf GCM} mode, and (4) generating a DSA key pair with a key length of only 512 bits. 

\subsubsection{Nonce reuse.} 
The ``Insecure Nonce Reuse" category, not mentioned in previous research, captures the reuse of the same nonce with the same key in symmetric encryption, which can compromise the strength of cryptography according to \cite{sp800-38c}. This category was detected in several programs in the LICMA dataset, such as {\sf PyNaClKeyDVA2.py}. However, LICMA merely labeled this issue as ``Use a hard-coded key".

\subsubsection{Use of insecure or deprecated cryptographic APIs}
The ``Use of deprecated OpenSSL functions and potential memory leak in OpenSSL lock initialization" category focuses on the use of {\sf \seqsplit{CRYPTO\_set\_id\_callback()}} and {\sf CRYPTO\_set\_locking\_callback()}, which have been deprecated since OpenSSL 1.1.0~\cite{ssl}. 

\subsubsection{Insecure cryptographic libraries}
The ``Insecure or deprecated cryptographic library usage" category highlights security risks associated with using insecure or deprecated cryptographic libraries, such as {\sf pyaes}\footnote{Available at https://github.com/
ricmoo/pyaes/commits/master}

. Although no conclusive evidence has been found regarding the insecurity of {\sf pyaes}, the fact that it has not been updated since 2017 suggests that it has likely been depreciated and increases its susceptibility to misuses and vulnerabilities.
	
\subsubsection{Misuse of tokens}
Three categories are related to the insecure use of tokens, which was rarely discussed in previous research.``Insecure usage of encryption tokens" describes attaching encryption tokens to URLs, an insecure practice not recommended by RFC 6750~\cite{rfc6750}. ``Insecure use of secrets for token generation" captures generating insecure tokens vulnerable to brute-force attacks, whereas NIST SP800-63~\cite{sp800-38b} recommends appropriate token lengths to mitigate this risk. The ``Insufficient token expiration and lack of proper error handling" category highlights the risks of using tokens for extended periods against NIST SP800-63's recommendations~\cite{digital_identity}. 
 
\subsubsection{Insecure certificate and server management}
Four categories fall in certificate and server management. ``Lack of public key pinning and certificate management issues" highlights the risk of servers without public key pinning, enabling attackers to use fraudulent certificates for man-in-the-middle attacks~\cite{chrome}. The ``Hardcoded Certificate" category points out that hardcoded certificates can be a single point of failure, as seen in certain versions of ALEOS~\cite{aleos} (CVE 2023-40464\footnote{Available
at https://nvd.nist.gov/vuln/detail/CVE-2023-40464}

). The ``Insecure cryptographic operation" category indicates the risks of long-lived certificates, which violate the recommendations of NIST SP800-52~\cite{tls}. Lastly, ``Short Certificate Validity Period” concerns using short-lived TLS certificates, which can lead to frequent renewals, management challenges, and potential service interruptions. NIST SP1800-16~\cite{ABC2020} provides guidance on meticulous management of TLS server certifications to avoid such risks.

\subsubsection{Misuse of new cryptographic elements}
Two new categories fall into this area. ``Improper cleanup of cryptographic operations and insecure Object Identifier (OID) creation" addresses the improper cleanup of cryptographic operations and the creation of OIDs without proper verification. The former violates NIST SP800-57 recommendations on erasing key usage traces after use \cite{sp800-57}, potentially exposing sensitive information to adversaries. Insecure OID creation, not mentioned in existing taxonomies, refers to practices that violate ISO/IEC 8825~\cite{iso8825} and RFC 5280~\cite{rfc5280} recommendations on OID creation, compromising OID uniqueness and leading to incorrect or exploitable cryptographic operations.

The ``Improper use of ENGINE API" category captures using OpenSSL's ENGINE API without first checking its availability. This oversight can cause implementation errors or unexpected behavior in encryption. According to the OpenSSL documentation~\cite{engine}, initialization to detect the availability of the ENGINE API is required before its usage. 
	
\subsubsection{New configuration errors}
As captured by ``Manipulation of {\sf crypto.policy} at runtime", modifying the {\sf crypto.policy} encryption policy at runtime can compromise security configurations and guarantees of the host systems. Official OpenSSL guidelines and documentation point out potential security risks, such as inconsistency and unpredictability in enforcing security policies across the application, associated with modifying cryptographic security levels and restrictions at runtime~\cite{Radack2010, OWASP21}.

\subsubsection{ Data signature}
As indicated by the ``Potential Misuse of Signature Function, Lack of Input Validation, and Improper Error Handling" category, the effectiveness of using {\sf crypto.Sign} to sign data depends on proper verification of its parameter settings, following guidelines such as RFC 3447~\cite{rfc3447} and the PyCryptodome library documentation~\cite{pycryptodome}. Without verifying the signature size or public key strength, the use of {\sf crypto.Sign} may lead to exploitation or bypass of the cryptographic signature.
	
\subsubsection{ Miscellaneous categories}
The rest 11 new CAM categories cover the following cases: (1) insecure practices not initializing {\sf OPENSSL\_init\_crypto()} properly or not verifying the return value of initialization~\cite{sslinit}; (2) configuring the SSH client to ignore the host SSH key verification using {\sf ssh.InsecureIgnoreHostKey()}, violating the recommendations of RFC 8446~\cite{rfc8446}; (3) implementing insecure secret sharing mechanisms, such as improper verification of transfer parameters~\cite{MKW20}; (4) using custom encoding schemes that lack data validation or security measures (as seen in CVE-2022-39955\footnote{Available at https://nvd.nist.
gov/vuln/detail/CVE-2022-39955}

); (5) defects in the OTP expiration verification logic violating the NIST SP800-63 recommendation~\cite{digital_identity};  (6) the lack of sufficient authentication measures when using functions like {\sf serialization.load\_pem\_private\_key()} and {\sf serialization.load\_ssh\_private\_key()} to deserialize keys, enabling the exploit of deserialization~\cite{deserialization}; (7) the use of cryptographic APIs related to inadequate error handling in public key decoding, (8) insecure handling and generation of secrets, (9) insecure private key storage and handling, (10) insecure data compression before encryption, and (11) insecure storage of encryption keys.

\subsection{RQ3: Usefulness}
As discussed above, CryMisTa demonstrated the ability to guide LLMs in generating more systematic and comprehensive taxonomies compared to existing approaches. Beyond aiding human developers in enhancing their knowledge and understanding of CAM, these taxonomies may also offer significant utility for designers of CAM detection tools. Specifically, the taxonomies can help cross-check CAM categories that are not yet covered by their tools, and the detailed explanations and code examples accompanying each CAM category provide valuable resources to inform the expansion of detection rules and engines within these tools.

To evaluate the potential application of the taxonomies generated by CryMisTa, we analyzed the 36 newly identified CAM categories in the GPT-4 taxonomy and summarized the following ways to derive detection rules from these categories: 
\begin{enumerate}
\item Use of insecure (often outdated or deprecated) cryptographic libraries, packages, and components.  Detection rules for these misuses can be defined as blacklisting the relevant entities. Detecting violations of such rules within the target code can be straightforwardly achieved by analyzing its import statements, code syntax, and dependencies.

\item Incorrect configuration and usage of API parameters. Several newly identified CAM categories are related to issues in the invocation order and parameter configuration of cryptographic APIs. These misuses often manifest as incorrect API usage (e.g., using incorrect APIs, invoking APIs in the wrong sequence, or omitting necessary APIs) or improperly configured parameters (e.g., incorrect settings of the operating mode, padding mode, initialization vector, and iteration count). Detection rules for such CAM categories can thus be formulated as temporal logic constraints specifying the correct orders of invoking cryptographic APIs and constraints specifying the correct parameter configurations for such APIs.   

\item Leakage of sensitive data. Variables and constants in code may store sensitive information, such as hardcoded certificates and tokens. Detection rules can hence be defined to restrict the encoding and usage of string- and integer-typed variables and constants to prevent sensitive data leakage or reuse.

\item Incorrect or lack of input validation and exception handling. For sensitive input data, rules can be defined to enforce proper validation of the format, type, range, or other attributes of the input before its usage. These rules should also include expected exception-handling logic when input data fails validation. Checking on these rules, however, are more complicated, due to the diversity of input data used by cryptographic APIs and the potential need for computationally expensive analytical techniques, e.g., taint analysis. 
\end{enumerate}

We selected 11 out of the 36 newly identified CAM categories and encoded them either as detection rules to enhance existing tools (e.g., CryScanner~\cite{CryScanner} and CogniCrypt$_{SAST}$~\cite{KNR+17}) or as straightforward detection logic (as Python scripts) to identify these types of CAM \footnote{Further details on how these CAM categories are leveraged to expand existing CAM detection capabilities are available in Appendix B and our GitHub repository.}. Our experiments demonstrated that all instances in $D$ that fall into these 11 categories were successfully detected by either the enhanced CryScanner and CogniCrypt$_{SAST}$ or by the encoded detection logic.  

\begin{figure}
\centering
\begin{lstlisting}[language=C, numbers=left, basicstyle=\ttfamily\footnotesize,numberstyle=\tiny, stepnumber=1, frame=single, caption={}, label={}]
OBJECTS
  a: OPENSSL_init_crypto()
  b: OPENSSL_cleanse()
  c: OPENSSL_add_all_digests()
ORDER
  a*b
CONSTRAINTS
FORBIDDEN
  c
\end{lstlisting}
\caption{Detection Rule for Inesure OpenSSL Initialization and Memory Handling in the CrySL Syntax}
\label{lst:rule}
\end{figure}

Consider the CAM category ``Lack of OpenSSL initialization and memory handling best practices” as an example. After reviewing its explanation and code example, we encoded it into a detection rule, as shown in Figure~\ref{lst:rule}, using the CrySL language~\cite{krugerCrySLExtensibleApproach2018}. This rule enforces using {\sf OpenSSL\_init\_crypto()} (instead of {\sf \seqsplit{OpenSSL\_add\_all\_digests()}}, listed in the {\tt FORBIDDEN} section) for OpenSSL initialization and {\sf OpenSSL\_cleanse()} for securely wiping memory, so as to ensure the confidentiality and integrity of encrypted data. By applying this rule to CryScanner~\cite{CryScanner}, it successfully detected this type of CAM in $D$ that it previously could not identify.

\section{Conclusion}\label{sec:conclu}
We have presented an approach leveraging the advanced capabilities of LLMs to automatically detect CAMs in realistic software code and classify them into a structured taxonomy. Our prototype, CryMisTa, has demonstrated the ability to work with multiple mainstream LLMs to generate more comprehensive CAM taxonomies from realistic code compared to existing approaches. CryMisTa also enabled the quantitative evaluation of LLM performance during the taxonomy construction. With more realistic code, CryMisTa can be used to guide LLMs in continuously refining and updating CAM taxonomies.

Future work includes: (1) fine-tuning general-purpose LLMs with the latest and more specific cryptographic information to improve their CAM detection capabilities; (2) tailoring the taxonomy construction prompts for different security analysis tasks to better align with specific use cases; and (3) overcoming the LLM token limits to analyze complex software projects.

\bibliographystyle{unsrt}  
\bibliography{ref0429, urls} 

\appendix
\section{ 
The Chain-of-Thought Prompts for The Detection of CAM Instances}\label{appendixA}
\begin{figure}[hbtp]
    \centering
    \includegraphics[width=0.45\textwidth]{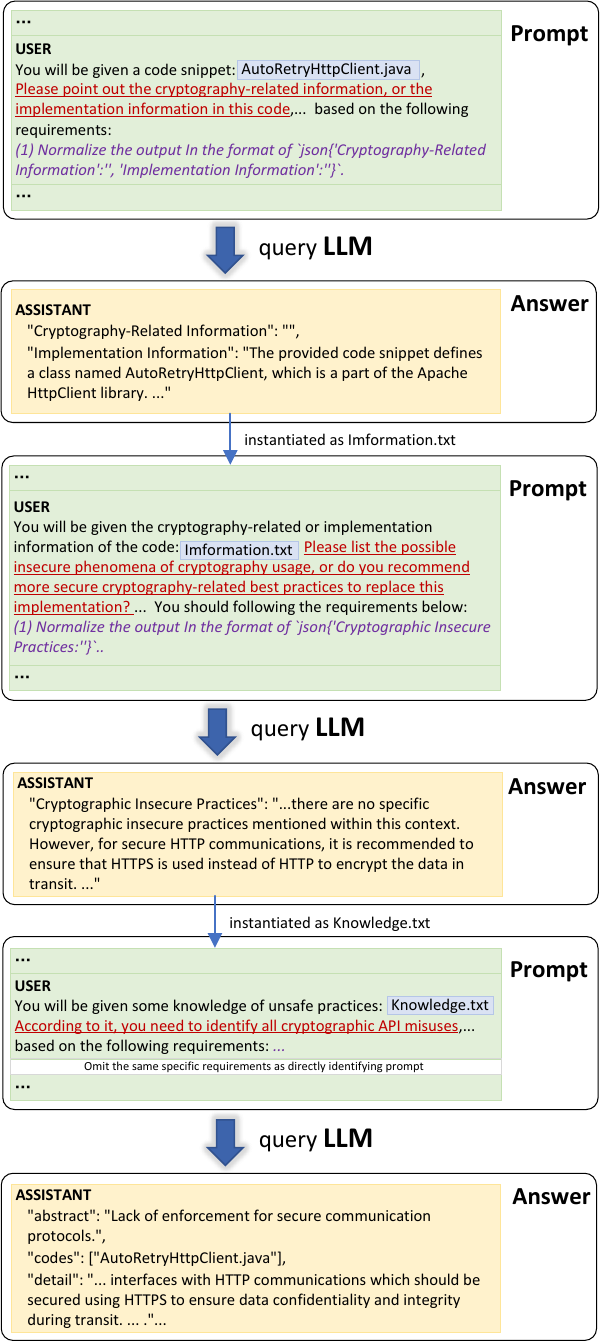}
    \caption{An example of CoT Identification}
    \label{fig:cot_id}
\end{figure}

Figure~\ref{fig:cot_id} demonstrates the Chain-of-Thought (CoT) promots implemented in CryMisTa to detect CAM instances in realistic code. To improve readability, Figure~\ref{fig:cot_id} also includes how GPT-4 responded to the CoT prompts to detect the ``Lack of enforcement for secure communication protocols" misuse in {\sf AutoRetryHttpClient.java} (available in our GitHub Repository). 

The CoT prompts first instruct the LLM to identify those elements of the input code related to cryptography (the top prompt in Figure~\ref{fig:cot_id}), then prompt the LLM to search its knowledge for related misuse practices (the middle prompt in Figure~\ref{fig:cot_id}), and lastly instruct it to apply this knowledge to analyze the input code (the bottom prompt in Figure~\ref{fig:cot_id}).

\newpage
\section{
Leveraging New CAM Categories to Expand CAM Detection Capabilities}\label{appendB}

Table~\ref{tab:misuse_rules} summarizes how we leveraged the selected 11 CAM categories to expand existing CAM detection capabilities. As shown in Table~\ref{tab:misuse_rules}, these rules are leveraged in one of the following three ways: (1) encoded as CrySL rules and applied to CryScanner and CongiCrypt$_{sast}$ tools (marked as ``Encoded as rules to expand CryScanner or CongiCrypt$_{sast}$" in the rightmost column of Table~\ref{tab:misuse_rules}); (2) being used to inform to expand or modify the detection engine of LICMA as LICMA is not a rule-based detection tool (marked as ``Update LICMA to align with the latest recommended iteration count."); and (3) being implemented as the Python detection logic (marked as ``Implemented as detection logic in Python").

\begin{table*}[!h]
\small
\caption{CAM Categories and Detection Expansion Methods}
\label{tab:misuse_rules}
\centering
\begin{tabular}{|p{3cm}|p{10cm}|p{3cm}|}
\hline
& \textbf{CAM Category} & \textbf{Means to Expand Detection Capabilities} \\
\hline

\multirow{9}{3cm}{Incorrect configuration and usage of API parameters} 
& \ \newline Title: Insufficient iterations count for PBKDF2. \newline
Explanation: Using a low number of iterations (1000) in the {\sf PBKDF2-HMAC} key derivation function compromises security by making the derived keys less resistant to brute-force attacks. Standards recommend a minimum of 10,000 iterations. 
& \ \newline Update LICMA to align with the latest recommended iteration count. \\
\cline{2-3}

& \ \newline Title: Insecure algorithm specified for SecretKeySpec and insecure key generation and management 

Explanation: Specifying the {\sf RC5} algorithm is insecure due to known vulnerabilities. Potential misuse in secret key management (e.g., static passwords) can lead to exploitable cryptographic keys. 
& \multirow{3}{3cm}{Encoded as rules to expand CogniCrypt$_{sast}$}. \\
\cline{2-2}

& \ \newline Title: Inappropriate GCM tag length 

Explanation: Using a GCM tag length shorter than 96 bits (NIST SP 800-38D standard) compromises data integrity and authenticity. Tag length must meet recommended standards. 
& \\ 
\cline{2-2}

& \ \newline Title: Inappropriate and unsupported IV and tag length for GCM 

Explanation: This misuse involves non-compliant IV and tag lengths in GCM encryption. IVs should be 96 bits, and tag lengths should meet or exceed 128 bits. Deviations weaken encryption security.
& \\ 
\cline{2-3}

& \ \newline Title: Insecure use of secrets for token generation 

Explanation: The token generation method used may not adhere to high security standards required by some applications, particularly concerning the token's length and complexity. 
& \multirow{3}{3cm}{\ \newline Implemented as detection logic in Python.}\\ 
\cline{2-2}

& \ \newline Title: Insecure nonce reuse 

Explanation: Reusing nonces with the same key in symmetric encryption leads to potential plaintext recovery or forgery attacks. Detected in files such as `PyNaClKeyDVA2.py`. Implemented as detection logic in Python.
& \\
\cline{2-2}

& \ \newline Title: Insecure OTP expiry validation 

Explanation: Flawed OTP expiry validation logic allows expired OTPs to be used, compromising time-bound authentication. Time checks must ensure OTPs are valid only within their intended lifespan.
& \\ 
\cline{2-3}

& \ \newline Title: Use of deprecated OpenSSL functions and potential memory leak in OpenSSL lock initialization 

Explanation: Deprecated OpenSSL functions (e.g., `CRYPTO\_set\_id\_callback`) may cause compatibility issues. {\sf init\_locks} lacks checks for successful memory allocation, risking null pointer dereference. 
&\multirow{2}{3cm}{\ \newline Encoded as rules to expand CryScanner.} \\
\cline{2-2}

& \ \newline Title: Lack of OpenSSL initialization and memory handling best practices 

Explanation: Failing to initialize OpenSSL or zeroize memory used for encryption can expose sensitive data. Best practices include using `OPENSSL\_init\_crypto` for initialization and `OPENSSL\_cleanse` for secure memory wiping. 
& \\ 
\hline

\ \newline Use of insecure cryptographic libraries, packages, and components
& \ \newline Title: Insecure or deprecated cryptographic library usage

Explanation: Deprecated libraries (e.g., `pyaes`) can introduce vulnerabilities. Secure cryptographic libraries like PyCryptodome are recommended. 
& \multirow{2}{3cm}{\ \newline Implemented as detection logic in Python.} \\
\cline{1-2}

\ \newline Leakage of sensitive data
& \ \newline Title: Hardcoded certificate

Explanation: Hardcoding certificates in application code increases the risk of compromise. Certificates should be stored externally to facilitate updates and revocation. 
& \\ 
\hline

\end{tabular}
\end{table*}

\end{document}